\documentstyle[prb,aps]{revtex}
\begin{document}
\preprint{NSF-ITP-97-114}
\draft

\title{\Large{Zener transitions between dissipative Bloch bands}}
 
\author{ Xian-Geng Zhao and W.-X. Yan$^*$}

\address{CCAST (World Laboratory) P.O. Box 8730, Beijing 100080, China\\
      Institute of Applied Physics and Computational Mathematics, 
     \\P.O. Box 8009, Beijing 100088, China\\
$^*$Institute of Theoretical Physics, 
        Academia Sinica, P.O. Box 2735, Beijing 100080, China\\} 

\author{Daniel W. Hone}

\address{Institute for Theoretical Physics, University of California, 
Santa Barbara, CA 93106 }

\date{\today}
\maketitle                          

\begin{abstract} 
Within a two-band tight binding model, we investigate the dynamics 
of electrons
 with Markoffian dephasing under the influence of static electric fields. 
With the help of both numerical and analytic
 calculations we find that the dephasing ultimately takes electrons 
which are initially  located in one 
miniband to equal population 
of the two minibands, instead of undergoing persistent Rabi flop, 
as they do in the absence of 
scattering.   Miniband localization is wholly destroyed 
by the intervention of dephasing.  We also obtain the effective 
decay time for the approach to equal band populations
 under conditions of small interband communication and in the long-time 
limit, through a perturbative calculation.  The decay rate shows 
characteristic sharp
peaks at values of the parameters which give Zener resonances. 

\end{abstract}
                   

\widetext

\section{Introduction}
Recently there has been intensive investigation$^{1-7}$ of the influence 
of external electric fields on semiconductor superlattices.
The availability of high-quality semiconductor superlattices, the possible 
practical application of new phenomena, and the surge of new interesting 
theoretical predictions have attracted both  theoretical
and experimental attention to this active field.
An important example is Bloch oscillation\cite{luban} (BO), one of
the earliest predictions of the band theory of solids.
Its observation in standard crystalline solids has been 
prevented\cite{phytoday} by
excessive scattering under all practical conditions.
But in semiconductor superlattices BOs have not only been observed
 experimentally\cite{expt} but have been further
 exploited to make  an extremely fast
 emitter of electromagnetic radiation --- the ``Bloch oscillator"\cite {phytoday}.
Many other new phenomena have been predicted 
theoretically and/or observed experimentally, including negative 
differential conductivity\cite{esaki,ndcond},
absolute negative conductance\cite{ancond}, inverse Bloch oscillation 
(strong THz-photocurrent resonance at the Bloch frequency)\cite{blochobs},
 dynamic localization\cite{blochobs,dunken,dunkenpla,shon,xg2,coul,holristow},
 band collapse\cite{holth,zak,honehol}, band suppression\cite{xg4bsup}, fractional 
Wannier-Stark Ladders\cite{xgjniu}, and
 multi-photon absorption\cite{multiph}.

From a theoretical point of view the simplest models involve only a 
single miniband.  But these miss all interesting interband phenomena, 
which are captured
most simply in a two band tight binding model first 
introduced\cite{fukuyama} by Fukuyama, Bari and Fogedby.  This model 
contains the essential physics of interband transitions, and 
can be compared in practice with realistic 
situations\cite{hnholt,honexg,xgniu2} where 
only a single pair of bands is important.
Rotvig, Jauho and Smith\cite{rjs,rjs2} have studied coherent 
transport of one-dimensional semiconductor superlattices within this model.
They found that coherent oscillation between
the minibands can occur at special values of the applied field, where
there are avoided crossings of the two interpenetrating 
Wannier-Stark Ladders (WSL) arising essentially
 from different bands\cite{rjs}. These are the so-called Zener resonances.
For electric field values between  Zener resonances there are 
stable plateaus, where the relative
population of the two bands is only weakly dependent on time.  In 
each plateau there are  small amplitude Bloch oscillations.  The 
number of these, $N$, corresponds to the $N$th Zener 
resonance\cite{rjs} (labelled with {\it decreasing} field, so that
the $N$th resonance arises from aligning adjacent WSL levels
in quantum wells $N$ lattice spacings apart).

In this paper, using the same model, we study the effects of scattering
from lattice imperfections on the dynamics of electrons, 
using a stochastic Liouville 
equation for the density matrix, with Markoffian dephasing.
By means of both analytic and numerical calculations we analyse the 
effects of scattering and miniband structure parameters on the 
time evolution of $\rho_-$, the difference in electron population of
the two minibands.
In contrast to previous work\cite{rjs,rjs2}, in which the
single quasimomentum component $\rho_-(k,t)$ was studied, 
we focus on the trace quantity
$\rho_-(t) = \sum_k\rho_-(k,t)$.
There are two reasons for this choice. The first is that the 
temporal evolution
of the sum over all quasimomenta is of direct physical interest, 
describing the dynamics of the full set of electrons.  The 
 individual $\rho_-(k,t)$ depend not only on time but on quasimomentum 
$k$, so we should integrate $\rho_-(k,t)$ over the whole Brillouin
 zone. The other is that, with scattering, $k$ is no longer a 
good quantum number --- even in a suitable transverse gauge (see
Ref.\ \onlinecite{honexg} and Eqs. (17) - (19) below), where $k$ 
{\it is} a good quantum number in the perfect lattice.  
We find that with no dissipation $\rho_-(t)$  periodically oscillates
between 1 and $-\rho_m$ (``Rabi flop" between the initial condition
of all electrons in one band and a fixed maximum exchange of a
fraction of that population with the other band: $0<\rho_m<1$) at avoided 
crossings, whereas when dephasing is introduced, the electrons 
initially assumed to be located in one 
miniband will ultimately be distributed equally between the two 
bands (band localization will be completely destroyed).

In section II, we present the model and derive the 
equation of motion in $k$ space. Section III provides numerical analysis of 
the effects of the single relaxation rate and of the miniband parameters  
on the Zener resonances.
Section IV contains the perturbative calculation of the density matrix and 
its limiting long-time
 behavior. We make some concluding remarks in section V.

\section{Model}

We consider the standard tight-binding model of a two-band system in 
a static electric field E. The model Hamiltonian\cite{fukuyama} can 
be written as
\begin{eqnarray}
 H = &&\sum_n\bigg[(\Delta_a + n\omega_B)a_n^\dagger a_n + (\Delta_b
+ n\omega_B)b_n^\dagger b_n \nonumber\\
&&- (W_a/4)(a^\dagger_{n+1} a_n + h.c.)
+ (W_b/4)(b^\dagger_{n+1} b_n + h.c.)\nonumber\\
&& + eER(a^\dagger_n b_n
+ b^\dagger_n a_n)\bigg].
\label{HAM}
\end{eqnarray}
Here the subscripts label the lattice sites and the lower and upper minibands
are designated by symbols $a$ and $b$, respectively.  We have introduced
the notation $\omega_B\equiv eEd$ for the Bloch frequency, which will appear
often below.  The first two terms
describe the site energies of the Wannier states in the presence of the
electric field, and $W_{a,b}$ are the widths of the isolated ($ E=0 $)
minibands induced by nearest neighbor hopping: $\epsilon^{a,b}(k) =
\Delta_{a,b} \mp (W_{a,b}/2) \cos (kd) $, where $d$ is the lattice constant.  
The last term is the on-site
electric dipole coupling between minibands; $eR$ is the corresponding dipole
moment.  This Hamiltonian does neglect Coulomb
interactions and electric dipole elements between Wannier states on different
sites, but it contains the essential physics for the 
problem\cite{fukuyama,honexg,rjs,rjs2}.
 Note that the
hopping parameters $W_{a,b}$ are written here with opposite signs, so that
with both parameters positive the band structure at $E=0$ is of the standard
nearly free electron character, with direct band gaps at the zone boundary.
But the calculation to follow is valid for arbitrary signs of the 
parameters.  

It is easily shown\cite{fukuyama,honexg} that the exact spectrum of $H$ 
is two interpenetrating Wannier-Stark Ladders.  But what do the 
corresponding states
represent in terms of the occupation of the original bands as a function
of time, and what is the influence of scattering?  For vanishing dipole
matrix element between bands, $R=0$, there is
no interband mixing.  Each of the two bands gives rise to a single WSL.
Clearly, when the electric field amplitude is such that the ladders 
become degenerate, even small values of $R$ lead to strong interband 
mixing.  The crossing of the ladders is ``avoided" by any finite
$R$, and the behavior at those avoided crossings (the Zener resonances)
is of particular importance and interest.

We start by defining the density matrix in the representation of the
two bands,

\begin{equation}
\rho(t)=\sum_{ijmn}\rho_{mn}^{ij}\xi_{m}^{i\dagger}\xi_{n}^{j},
\end{equation}
 where $i,j = 1$ or $2$ are band indices: $\xi_{m}^{1\dagger}$ 
($\xi_{m}^{1}$) and $\xi_{m}^{2\dagger}$
 ($\xi_{m}^{2}$) designate $a_m^{\dagger}$ ($a_m$) and $b_m^\dagger$ 
($b_m$), respectively.
 Within a constant relaxation rate approximation\cite{dunlap}, the density
 matrix $\rho(t)$ satisfies the following stochastic Liouville equation (SLE)
(we set $\hbar=1$ throughout this paper),
\begin{equation}
i\frac{d\rho}{dt}=[H, \rho(t)]-i\Gamma \rho(t).
\end{equation}
Here $\Gamma \rho$ describes the relaxation 
of the off-diagonal elements of $\rho$ through dephasing: 
\begin{equation}
\Gamma \rho=\sum_{ijmn}(1-\delta_{ij}\delta_{mn})
\Bigl(\alpha_{ij}\xi_{m}^{i\dagger}\xi_{n}^{j}\Bigr).
\end{equation}
The utility of this simplest form of the SLE has been discussed
by Kenkre and collaborators (see Ref. \onlinecite{dunlap} and
references therein).  The parameters $\alpha_{ij}$ measure the loss
of phase coherence between sites, or the scattering lifetime of
band states labeled by quasimomentum.  The relaxation to zero of
the lattice site off-diagonal elements of $\rho$ is then effectively
an infinite temperature approximation.  The finite temperature
corrections are essential\cite{dunlap,alekseev} 
to understanding the current response
to an electric field, a response which vanishes in the absence of
a difference in thermal equilibrium population between states of
different energy, but those corrections should not play a major
role in the phenomena studied here.  We note that in this limit
there is no need to distinguish\cite{alekseev} between the
relaxation rates of energy and momentum. 

Since we are interested in the dynamics of occupation of 
various band states, it is convenient to work in a wave vector
basis, by Fourier transforming the density matrix.  In general, 
since $\rho_{mn}$ is not translationally invariant (a function
only of $m-n$), we have a full set $\rho^{ij}_{kq} = \sum_{mn}
\rho_{mn}^{ij}(t)\exp[-ikm+iqn]$ of Fourier components.  But we
will be interested in the wave vector diagonal band occupation
numbers $\rho^{ij}_{kk}(t) \equiv \rho^{ij}(k,t)$.  These evolve
according to the corresponding Fourier transform of the SLE:

\begin{eqnarray}
i\frac{\partial}{\partial t} \rho^{11}(k,t)=&& i\omega_B\frac{\partial}
{\partial k}
\rho^{11}(k,t)-eER[\rho^{12}(k,t)-\rho^{21}(k,t)]\nonumber\\
&&-i\alpha_{11}\rho^{11}(k,t)
+i\alpha_{11}\int_0^{2\pi}\frac{dk'}{2\pi}~\rho^{11}(k',t)~~,
\end{eqnarray}
\begin{eqnarray}
i\frac{\partial}{\partial t} \rho^{22}(k,t)=&& i\omega_B\frac{\partial}
{\partial k}
\rho^{22}(k,t)+eER[\rho^{12}(k,t)-\rho^{21}(k,t)]\nonumber\\
&&-i\alpha_{22}\rho^{22}(k,t)
+i\alpha_{22}\int_0^{2\pi}\frac{dk'}{2\pi}~\rho^{22}(k',t)~~,
\end{eqnarray}
\begin{eqnarray}
i\frac{\partial}{\partial t} \rho^{12}(k,t)=&& i\omega_B\frac{\partial}
{\partial k}
\rho^{12}(k,t)-eER[\rho^{11}(k,t)-\rho^{22}(k,t)]\nonumber\\
&&+\Bigl(\Delta-W\cos k\Bigr) \rho^{12}(k,t)
-i\alpha_{12}\rho^{12}(k,t)~~,
\end{eqnarray}
\begin{eqnarray}
i\frac{\partial}{\partial t} \rho^{21}(k,t)=&& i\omega_B\frac{\partial}
{\partial k}
\rho^{21}(k,t)+eER[\rho^{11}(k,t)-\rho^{22}(k,t)]\nonumber\\
&&-\Bigl(\Delta-W\cos k\Bigr) \rho^{21}(k,t)
 -i\alpha_{21}\rho^{21}(k,t)~~,
\end{eqnarray}
where we have adopted the notation $\Delta\equiv\Delta_a - \Delta_b$
and $W \equiv (W_a+W_b)/2$.  By introducing

\begin{eqnarray}
\rho_+(k,t)=\rho^{11}(k,t)+\rho^{22}(k,t)~,\\
\rho_-(k,t)=\rho^{11}(k,t)-\rho^{22}(k,t)~,\\
\rho_{+-}(k,t)=\rho^{12}(k,t)+\rho^{21}(k,t)~,\\
\rho_{-+}(k,t)=i[\rho^{21}(k,t)-\rho^{12}(k,t)]~,
\end{eqnarray}
and, for simplicity, taking $\alpha_{11}=\alpha_{22}=\alpha_1,~~
\alpha_{12}=\alpha_{21}=\alpha_2$ to reduce the number of parameters
in the theory, we obtain

\begin{equation}
\frac{\partial}{\partial t}\rho_+(k,t)-\omega_B\frac{\partial}{\partial k}\rho_+(k,t)=
-\alpha_1\left[\rho_+(k,t)-\int_0^{2\pi}\frac{dk'}{2\pi}~\rho_+(k',t)\right]~,
\end{equation}
\begin{equation}
\frac{\partial}{\partial t}\rho_-(k,t)-\omega_B\frac{\partial}{\partial k}\rho_-(k,t)=
-2eER \rho_{-+}(k,t)-\alpha_1\left[\rho_-(k,t)-\int_0^{2\pi}\frac{dk'}
{2\pi}~\rho_-(k',t)\right]~,
\end{equation}
\begin{equation}
\frac{\partial}{\partial t}\rho_{+-}(k,t)-\omega_B\frac{\partial}
{\partial k} \rho_{+-}(k,t)=
\Bigl(\Delta - W\cos k\Bigr) \rho_{-+}(k,t)
 -\alpha_2\rho_{+-}(k,t)~,
\end{equation}
\begin{equation}
\frac{\partial}{\partial t}\rho_{-+}(k,t)-\omega_B\frac{\partial} 
{\partial k} \rho_{-+}(k,t)=
-\Bigl(\Delta - W\cos k\Bigr) \rho_{+-}(k,t)
 +2eER \rho_-(k,t)-\alpha_2\rho_{-+}(k,t)~.
\end{equation}

The equation for $\rho_+(k,t)$ describes particle conservation.  It is
decoupled from the others, and we will
ignore it in the following discussion. The equations for 
$\rho_-(k,t),~~\rho_{+-}(k,t),$
and $\rho_{-+}(k,t)$ can be reduced to the following ordinary differential 
equations in an accelerated basis\cite{kria}, $k(t) = k-\omega_B t$ or,
equivalently, in the transverse or vector gauge discussed above and in
Ref.\ \onlinecite{honexg},
\begin{equation}
\frac{d}{dt}X(k,t)=-2eER~Z(k,t)-\alpha_1[X(k,t)-X_0(t)],
\end{equation}
\begin{equation}
\frac{d}{dt}Y(k,t)= [\Delta - W\cos(k-\omega_B t)]Z(k,t) 
 -\alpha_2Y(k,t),
\end{equation}
\begin{equation}
\frac{d}{dt}Z(k,t)= -[\Delta - W\cos(k-\omega_B t)]Y(k,t) 
+2eER~X(k,t)-\alpha_2Z(k,t)~.
\end{equation}

Here $X(k,t)=\rho_-(k-\omega_B t,t)$, $Y(k,t)=\rho_{+-}(k-\omega_B t,t)$,
$Z(k,t)=\rho_{-+}(k-\omega_B t,t)$, and
$X_0(t)=\int_0^{2\pi}X(k,t)dk/2\pi=\rho_-(t)$,~~.
In the absence of any scheme for solving these coupled equations 
analytically, we turn to numerical solutions.

\section{Zener Resonance With Dissipation: Numerical Results}

In this section we focus on the influence of dissipation on
$\rho_-(t)$, the difference in population between the two
minibands.  For further simplicity the  relaxation rates 
$\alpha_1$ and $\alpha_2$ 
have been set equal, $\alpha_1 = \alpha_2 = \alpha$, in the following.

We emphasize again that due to scattering, the time dependences 
of $X(k,t), Y(k,t)$ and $Z(k,t)$ as described in Eqs. (17) - (19) 
depend on the sum of these quantities 
over {\em all} values of the wave vector, whereas in the absence of 
scattering, each value of $k$ is independent\cite{rjs,rjs2}.
As we have discussed above, we are particularly interested in the 
dynamics of the difference in electron population between 
the two minibands\cite{rabif}, as described by $\rho_-(t)$.
In the perfect lattice some one-component diagonal terms $X(k,t)$
undergo Rabi oscillation\cite{xgniu2,rjs,rjs2} --- i.e., $X(k,t)$ oscillates 
between $+1$ and $-1$ with a
period approximately determined by the value given in Ref.  
\onlinecite{xgniu2}, if we choose
the electric field at a value corresponding to one of the avoided crossings
 of the two interpenetrating Wannier-Stark Ladders.
 But further analysis shows that for some other wave vectors $X(k,t)$ will 
{\em not} undergo complete Rabi oscillations (will not reach the full 
extreme values of $\pm 1$), even if we choose
 the electric field to be at the avoided crossings.  
This is depicted in the lower panels of both Figs. 1 and 2, which differ 
only by the choice of system parameters, listed in the figure captions. 
The upper panel in each corresponds to $k=0$ and the lower panel to $k=\pi$.
It can be clearly seen  that 
$X(k=\pi,t)$ falls short of the minimum value of -1 in both cases. 
But, for the parameters of Fig. 2, the minimum is more closely 
approached than it is in Fig. 1,  although in Fig. 2 interband 
communication has been decreased.  This can be understood by making 
the coordinate transformation $t=t'+k/\omega_B$ in 
Eqs. (17) - (19) ( with no relaxation: $\alpha = 0$).
Then all quantities depend only on the single parameter $t'$:
$X(k,t)=X(t'), Y(k,t)=Y(t'), Z(k,t)=Z(t')$.   The initial
conditions are set at a time $t$ and therefore are still $k$ dependent
in terms of $t'$.  But  
when $E$ is very large, we have $t\approx t'$, so that $X(k,t)$, 
$Y(k,t)$, and $Z(k,t)$ will be independent of $k$.  Then all $X(k,t)$
 will approximately undergo full Rabi oscillations.

In general this is the dominant reason that $\rho_-(t)$, which represents
a sum over all wave vectors, oscillates between certain negative values 
 (but never reaches $-1$ ) and $+1$, no matter how we choose the 
electric field.  We demonstrate this
in Fig. 3a, where we use the same parameters as those of Fig. 1.
In this figure, Zener tunneling and the more rapid BO can both still 
be clearly
 identified, but some oscillations in the stable plateaus have been
 smeared out. In particular, when $\rho_-(t)$ reaches $+1$, only a single 
sharp cusp has survived. The smearing is due to interference between 
the BO's for different values of $k$, which appear in slightly 
different positions, as can be
 seen by comparing the upper and lower panels of Figs. 1 and  2.  This 
interference is also a factor (but not the main one) in suppressing the
minimum of $\rho_-(t)$ from $-1$ at Zener resonances.

With the introduction of dissipation the dynamics change dramatically 
(even in the regime where the scattering rate is less than the 
Bloch frequency, $\alpha/\omega_B < 1$, so that some
BO's remain well defined; we will treat only this regime\cite{fn}).    
We demonstrate this in Fig. 3b, where we use the same parameters
as in Fig. 3a except that $\alpha/\omega_B=0.02$ instead of zero.  
In this figure, we see that 
$\rho_-(t)$ still oscillates
 between negative and positive values, but
 the oscillation amplitude decreases strikingly with increasing time.  
Plateaus can still be identified in several earlier peaks
 of the envelopes, but 
the BO's are increasingly suppressed in successive plateaus.
If we choose the ratio  $\alpha/\omega_B$ larger ( e.g., $\alpha/\omega_B=
0.3$ in Fig. 3c, where we use the same other parameters as in Fig. 3a), 
a few BO's can be seen but $\rho_-(t)$ never becomes negative. 
This demonstrates that observation of Rabi oscillations is in general 
more difficult than that of BO.

In the absence of scattering we can localize the 
electrons in one of the minibands simply by choosing the electric field 
amplitude to be well away from
 the avoided crossings.  This can be seen
through numerical calculation of  $\rho_-(t)$.
However, the introduction of dephasing destroys this picture completely. 
For {\em any} value of the electric field the electron density will be
distributed equally between the two minibands at long times. This can be 
seen clearly by comparing
Figs. 4a and 4b, where we use the same  superlattice parameters
 as in Fig. 2, and the BO frequency  $\omega_B$ has been changed from 
the value at an avoided crossing  of 10.2 meV to $9$ meV.
In Fig. 4a there is no dissipation, while in Fig. 4b,  
$\alpha/\omega_B=0.35$.

\section{Perturbation Theory and Long Time Transitions:  Exact Results}

Under certain conditions perturbative calculations
can give some useful results.  
Setting $A(k,t)=Y(k,t)+iZ(k,t),~ B(k,t)=Y(k,t)-iZ(k,t), 
\mu \equiv 2eER,$  and recalling $\alpha_1=\alpha_2=\alpha$, 
we can rewrite the Eqs. (18) and (19) as
\begin{equation}
\frac{d}{dt}A(k,t)=-i[\Delta-W\cos(k-\omega_B t)]A(k,t)-\alpha A(k,t)
+i\mu X(k,t),
\end{equation}
\begin{equation}
\frac{d}{dt}B(k,t)=i[\Delta-W\cos(k-\omega_B t)]B(k,t)-\alpha B(k,t)
-i\mu X(k,t).
\end{equation}
In order to facilitate the following perturbative manipulation, we make the 
transformations:
\begin{equation}
A(k,t)=a(k,t)e^{-\alpha t-i\int^{t}_{0} dt'[\Delta-W\cos(k-\omega_B t')]},
\end{equation}
and
\begin{equation}
B(k,t)=b(k,t)e^{-\alpha t+i\int^{t}_{0} dt'[\Delta-W\cos(k-\omega_B t')]}.
\end{equation}
Then Eqs. (17) - (19) can be written equivalently as
\begin{equation}
\frac{d}{dt}\left[ e^{\alpha t}X(k,t)\right]=
-\frac{\mu}{2i}[a(k,t)e^{-i\int^{t}_{0} dt'[\Delta-W\cos(k-\omega_B t')]}-b(k,t)e^{i\int^{t}_{0} dt'[\Delta-W\cos(k-\omega_B t')]}]
+\alpha X_0(t)~,
\end{equation}

\begin{equation}
\frac{d}{dt}a(k,t)=i\mu X(k,t)e^{\alpha t+i\int^{t}_{0}
dt'[\Delta-W\cos(k-\omega_B t')]}~,
\end{equation}

\begin{equation}
\frac{d}{dt}b(k,t)=-i\mu X(k,t)e^{\alpha t-i\int^{t}_{0}
dt'[\Delta-W\cos(k-\omega_B t')]}~.
\end{equation}
 As usual, we take the electrons to be initially located in band a. 
If the communication between the two bands is small,
 we can perform perturbative calculations by integrating 
 Eqs. (25) and (26) and substituting the resultant expressions
for $a(k,t)$ and $b(k,t)$ in terms of $X(k,t)$ into (24), which 
is then integrated by iteration to the first nontrivial order.  
 This gives, after summation over $k$, the approximate integral equation,
\begin{eqnarray}
\rho_-(t)=&&X_0(t)=
e^{-\alpha t}\Bigl\{1 +\alpha\int^{t}_{0}dt'X_0(t')e^{\alpha t'}
\nonumber\\
&& - \mu^2\int^t_0dt'\int^{t'}_{0}dt''[1+\alpha\int^{t''}_{0}d\tau X_0(\tau)
e^{\alpha \tau}]J_0\Bigl ( \frac{2W}{\omega_B}\sin\frac{\omega_B(t'-t'')}{2} \Bigr ) \cos
\Delta(t'-t'') \Bigr\}~,
\end{eqnarray}
 where $J_0$ is the ordinary Bessel function of order zero.
Using the identity
\begin{equation}
J_0(a\sin x) = \sum_m J_m^2(a/2) e^{2imx},
\end{equation}
we can readily Laplace transform Eq. (27): 
\begin{equation}
X_0(P)=\frac{1-\mu^2f(P+\alpha)}{P+\alpha\mu^2f(P+\alpha)}~,
\end{equation}
where $f(P+\alpha)$ is defined by 
\begin{equation}
f(P+\alpha)=\frac{J_0^2(W/\omega_B)}{(P+\alpha)^2+\Delta^2}+
\sum\limits^\infty_{m=1}J_m^2(\frac{W}{\omega_B})\Bigl[\frac{1}
{(P+\alpha)^2+(m\omega_B+\Delta)^2}+\frac{1}
{(P+\alpha)^2+(m\omega_B-\Delta)^2}\Bigr]~.
\end{equation}
Then (29) gives the long time behavior of $X_0(t)=\rho_-(t)$ as
\begin{equation}
\rho_-(t\rightarrow\infty)=\Bigl[ 1-\mu^2f(\alpha)\Bigr]e^{-t/\tau}~,
\end{equation}
where $\tau$, the effective decay time, is defined as $\tau^{-1}=
\alpha\mu^2f(\alpha)$.
From Eqs. (30) and (31) and the definition of effective scattering 
time $\tau$,
 setting $\tau_0=\alpha^{-1}$, we find
\begin{equation}
\frac{\tau^{-1}}{\tau_0^{-1}}=(\mu\tau_0)^2\left\{
\frac{J_0^2(W/\omega_B)}{1+(\Delta\tau_0)^2}+
\sum\limits^{\infty}_{m=1}
J_m^2(\frac{W}{\omega_B})\left[ 
\frac{1}{1+(\Delta/\omega_B +m)^2(\omega_B\tau_0)^2}
+\frac{1}{1+(\Delta /\omega_B-m)^2(\omega_B\tau_0)^2}\right]
\right\}~.
\end{equation}
This relation describes the leading long time dynamics of 
electrons.
The structure of Eq. (32) shows that when $\Delta/\omega_B$ is an integer,
 ${\tau^{-1}}/{\tau_0}^{-1}$ will reach a local maximum;  this is 
illustrated in Fig.~5. 
 These resonant peaks indicate that the excited state (i.e., the rungs 
of the upper one of the two interpenetrating Wannier-Stark Ladders) 
in one well is degenerate with the ground state of a well m unit 
cells away.  When the scattering rate $\alpha = \tau_0^{-1}$ becomes 
smaller, the peaks in $\tau^{-1}/{\tau_0^{-1}}$ become 
higher and sharper (half width in $\Delta/\omega_B$ of order
$(\omega_B\tau_0)^{-1}=\alpha/\omega_B$), as one expects intuitively.  
Note that the decay 
time $\tau$ in the figure is always longer than
the scattering time $\tau_0$, even at the peaks, which was tacitly 
assumed in setting $f(P+\alpha) \approx f(\alpha)$ at the dominant
(smallest negative $P$) pole in (29) above.

\section{Concluding Remarks}

We have discussed the dynamics of electrons in a two-band superlattice
under the influence of static electric fields, with Markoffian dephasing. 
We  have given both analytic and 
numerical analyses of the effects of scattering and miniband structure 
on the Zener resonances and miniband localization.
 We found that dephasing will ultimately force the electrons 
initially located in one miniband to populate equally the two
 minibands.  Under conditions of  small interband coupling, 
and in the long time limit, 
we obtain the effective decay rate, whose sharp peaks are
a signature of Zener resonances.
 
Experimentally the phenomena predicted in this paper are accessible
in semiconductor superlattices with wells of dimensions on the order
of 100 Angstroms and electric fields of a few kilovolts per cm. 
 This superlattice can be grown dimerized, with the unit cell consisting,
 e.g., of a pair of quantum wells, separated from the adjacent pair 
by a wider barrier than that which separates the two wells of the 
basis pair. Then the lowest level of an individual well is split
within the basis pair. These two levels form a pair of minibands in
the superlattice which is well separated energetically from higher
levels, and all but that pair can reasonably be neglected in  
studying the dominant electron dynamics\cite{hnholt,honexg}.
Then we can probe radiation induced by the oscillating dipole 
associated with Rabi flop in this system\cite{xgniu2}.

Work is under way to extend these results to understand the influence
of scattering on the behavior of a semiconductor superlattice in 
the presence of strong time periodic, as well as static, electric
fields.  Of particular practical interest are superlattices
subjected to the terahertz radiation of a free electron laser.
A study of the single miniband case, including a discussion of 
chaotic dynamics in the presence of dissipation, has recently been
published\cite{alekseev} by Alekseev, et al.

\begin{center}

 {\bf ACKNOWLEDGMENT}

\end{center}
The authors thank Prof. W.-M. Zheng for useful and stimulating discussions.
This work was supported in part by the National Natural Science Foundations 
of China and a grant of the China Academy of Engineering and Physics, and
in part by the U.S. National Science Foundation Grant PHY94-07194.

\newpage

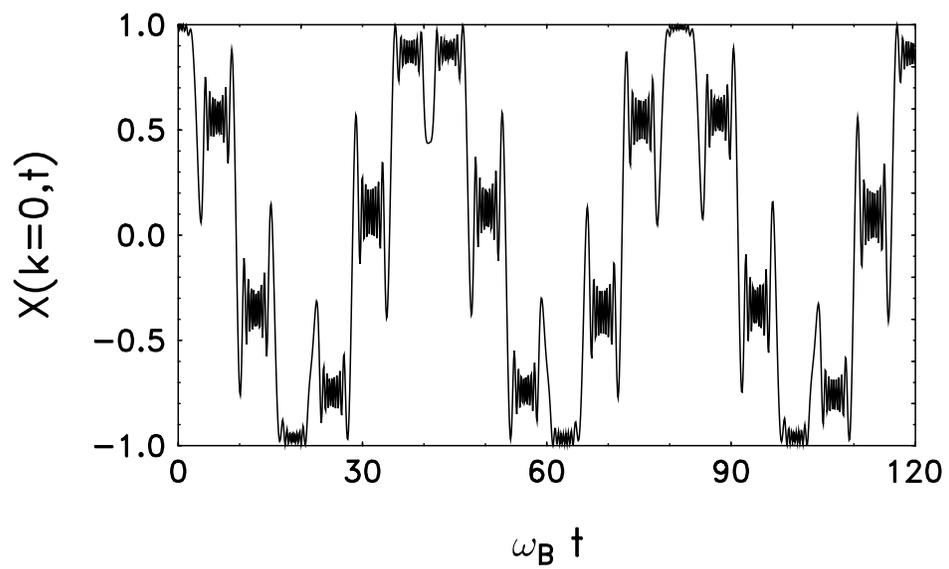
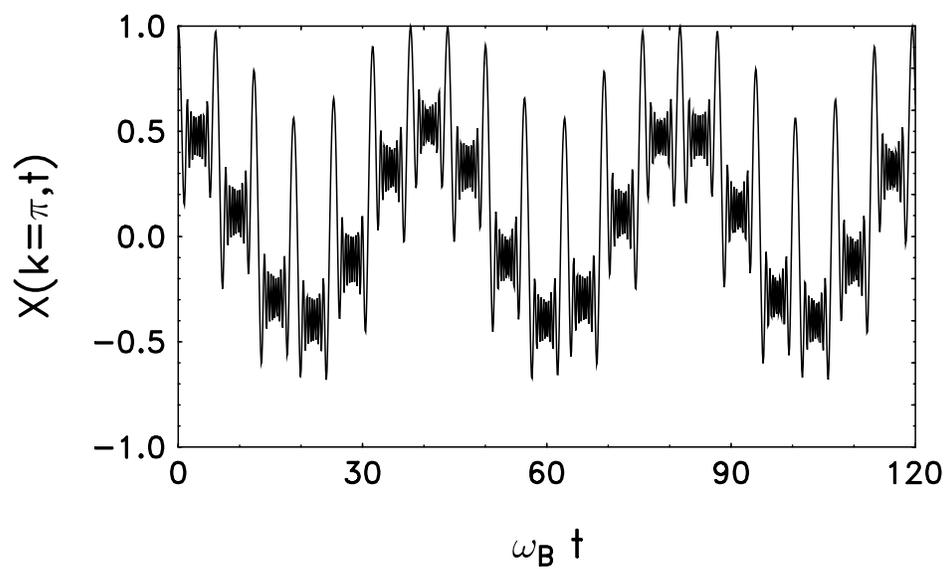
\begin{figure}
\caption{Dependence on wave vector $k$ of both oscillation 
amplitudes and the position
 of BO of a single component $X(k,t)$, plotted as functions of 
dimensionless time, $\omega_B t$. In the upper panel (with $k=0$), full
Rabi oscillations can be clearly seen, while in the lower panel (with $k=\pi$),
they are incomplete. In both figures
 $W=(W_a+W_b)/2$=18 meV, $\Delta=\Delta_a-\Delta_b=20$ meV, $eER=2.09$ meV, and 
$\omega_B=2.32$ meV is at an avoided crossing of the interpenetrating WSL from 
the two bands.}
\end{figure}

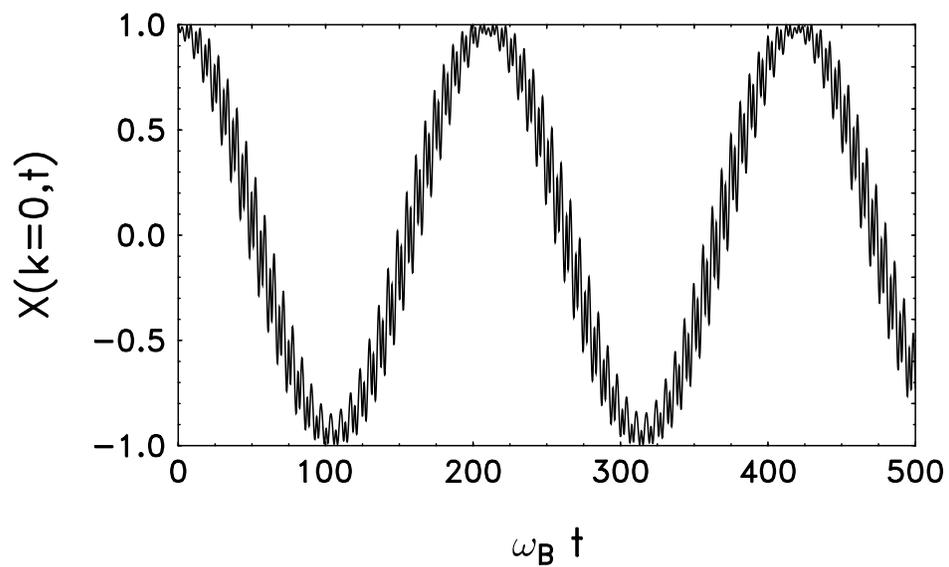
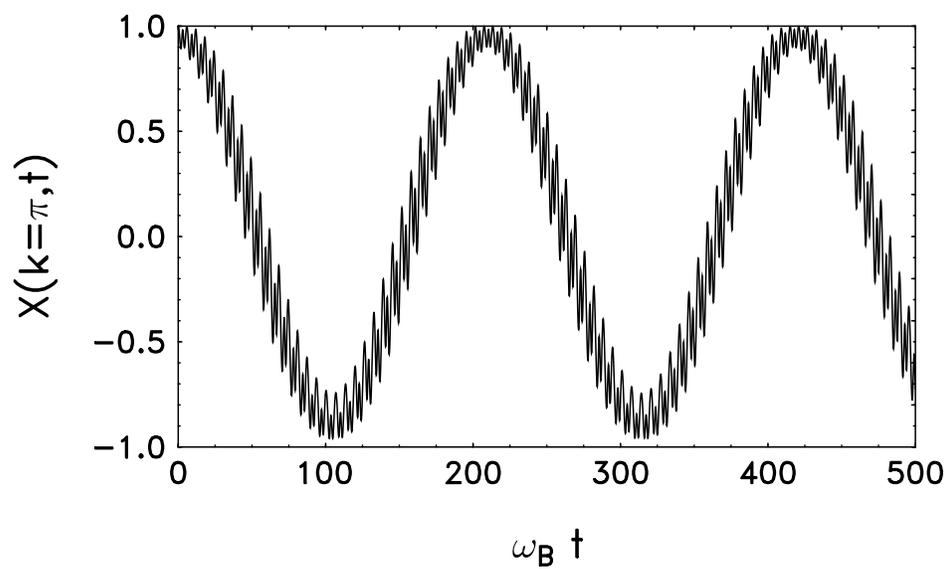
\begin{figure}
\caption{The same as in Fig. 1,  with  $W=8.6$ meV,
  $\Delta=20$ meV, $eER=1.84$ meV, and $\omega_B$ has been increased 
to $10.2$ meV (still at an avoided crossing).}
\end{figure}

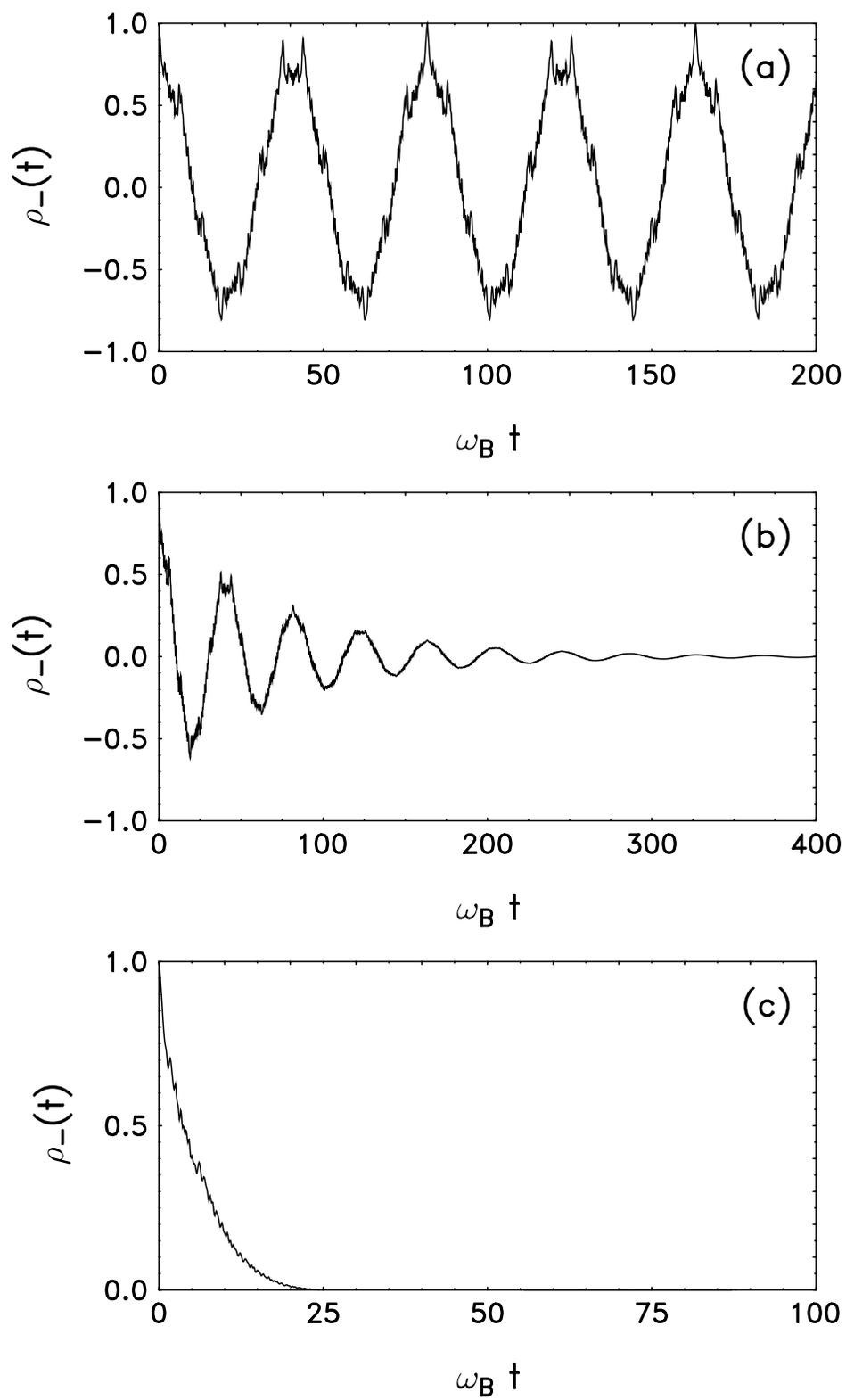
\begin{figure}
\caption {The density matrix $\rho_-(t)$, with and without dissipation.
We use the same superlattice
 parameters as those of Fig. 1.
(a) no scattering, $\alpha/\omega_B=0$;
(b) small scattering rate, $\alpha/\omega_B = 0.02$;
(c) large scattering rate, $\alpha/\omega_B = 0.3$.}
\end{figure}

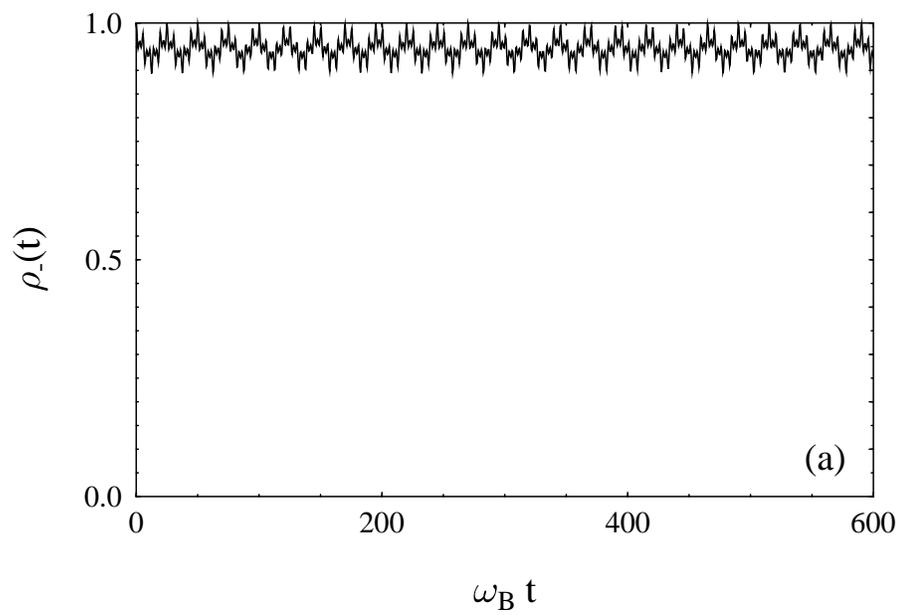
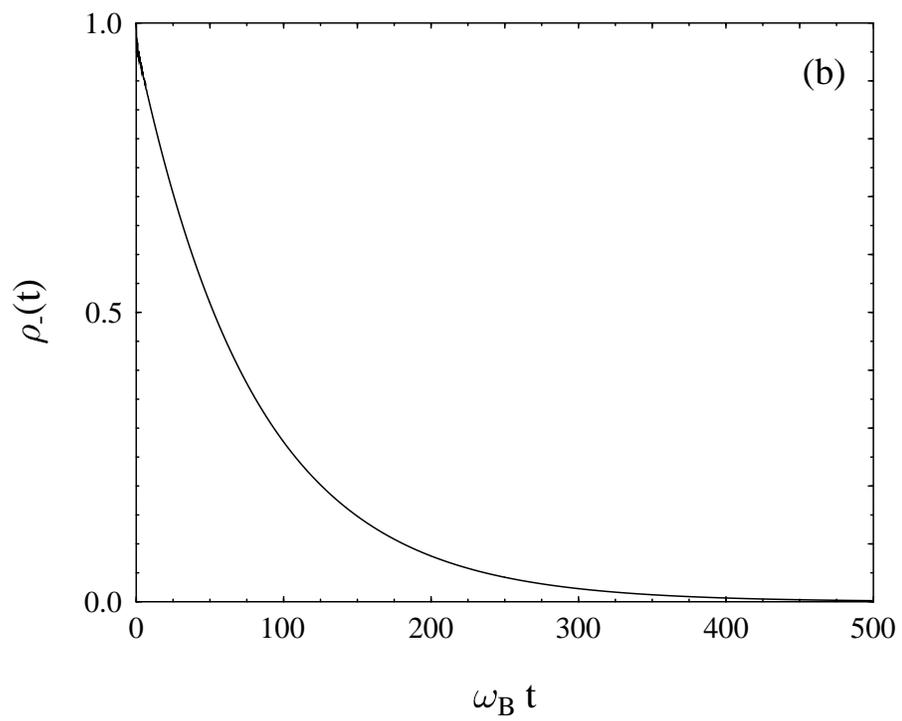
\begin{figure}
\caption {The effect of scattering on miniband localization.  
 (a)  no scattering; 
 (b) the complete destruction of localization by dephasing
(with $\alpha/\omega_B=0.35$).
Here $W=(W_a+W_b)/2$=8.6 meV, $\Delta=\Delta_a-\Delta_b=20$ meV,
 $eER=1.84$ meV,  and  $\omega_B = 9$ meV.}
\end{figure}

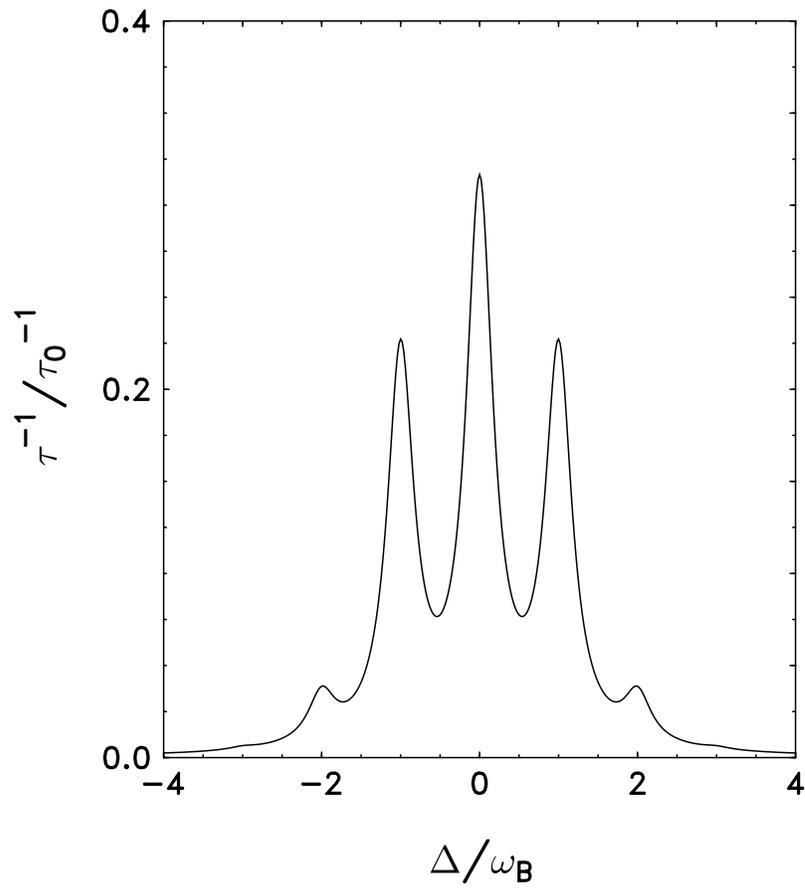
\begin{figure}
\caption {Long time decay rate $\tau^{-1}/\tau_0^{-1}$ of $\rho_-(t)$ 
(see Eqs. (31) and (32)).
The local maxima appear when $\Delta/\omega_B$ is an integer.
Here $\mu/\omega_B=0.18$, $\alpha/\omega_B=0.2$, and 
$W/\omega_B=1.3$.}
\end{figure}


\begin{references}
\bibitem{fn} The very observation of BO, and measurements\cite{duffield} 
of cyclotron resonance linewidths suggest as reasonable values
for $\alpha$ in practical semiconductor superlattices something on
of the order of a terahertz.  Then for a typical lattice constant $d$ of  
100 Angstroms the Bloch frequency $\omega_B = eEd$ becomes equal to
$\alpha$ for electric fields $E$ of order 1 kV/cm.
\end{references}
\end{document}